\documentclass[11pt,a4paper]{article}
\pdfoutput=1
\usepackage{jheppub}

\usepackage[utf8]{inputenc}
\usepackage{graphicx}
\usepackage{amsmath,amsfonts,amssymb}
\usepackage{epsfig,color}
\usepackage[outdir=./]{epstopdf}
\usepackage[T1]{fontenc}
\usepackage{graphicx}% Include figure files
\usepackage{color} % use colored fonts
\usepackage{dcolumn}% Align table columns on decimal point
\usepackage{bm}% bold math
\usepackage{epsfig}
\usepackage[subfigure]{graphfig}
\usepackage{latexsym}
\usepackage{amsmath}
\usepackage{amsfonts}
\usepackage{amsxtra}
\usepackage{hyperref}
\usepackage{placeins}

\newcommand{\lwrsim}{\raise0.3ex\hbox{$<$\kern-0.75em\raise-1.1ex\hbox{$\sim$}}}
\def\krto{ {\,\,\lower .8ex\hbox {$\longrightarrow \atop k \rightarrow 0$}\,\,}}

\def\bea{\begin{eqnarray} }
\def\beq{\begin{eqnarray} }

\def\eea{\end{eqnarray}}
\def\eeq{\end{eqnarray}}

\newcommand{\nn}{\nonumber \\ }

\preprint{\begin{tabular}{r}JLAB-THY--21-3520
\end{tabular}}
\begin{document}

%%%%%%%%%%%%%%%%%%%%%%%%%%%%%%%%%%%%%%%%%%%%%%
%% title page 
%%%%%%%%%%%%%%%%%%%%%%%%%%%%%%%%%%%%%%%%%%%%%%

\title{One-loop structure of  parton distribution for the gluon condensate  and ``zero modes''}

\author[a,b]{Anatoly Radyushkin}
\author[a,b]{and Shuai Zhao}

\affiliation[a]{Old Dominion University,  \\4600 Elkhorn Ave., Norfolk, VA 23529, USA}
\affiliation[b]{Thomas Jefferson National Accelerator Facility,  \\ 12000 Jefferson Ave., Newport News, VA 23606, USA}

\emailAdd{radyush@jlab.org}
\emailAdd{szhao@odu.edu}

\abstract{
We present  results for  one-loop corrections to 
the recently introduced ``gluon condensate'' PDF $F(x)$.
In particular, we give expression for the   $gg$-part of   its evolution kernel. 
To enforce  strict compliance with the gauge invariance requirements,
we have used  on-shell states for external gluons, and have obtained identical results 
both in Feynman and light-cone gauges. 
No ``zero mode'' 
 $\delta (x)$  terms were found for the  twist-4 gluon PDF $F(x)$. 
However a 
 $q^2 \delta (x)$ term was found for   the $\xi=0$  GPD $F(x,q^2)$ at nonzero  momentum transfer $q$. 
Overall, our results do not agree with the original  attempt of  one-loop calculations  of  $F(x)$ 
for gluon states, 
which 
  sets  alarm warning for calculations that use matrix elements   with 
 virtual external  gluons and  for lattice  renormalization procedures 
based on their results.

}
  
\maketitle

\date{\today}

\flushbottom
%%%%%%%%%%%%%%%%%%%%%%%%%%%%%%%%%%%%%%%%%%%%%%%%%%%%%%%%%
%% body of the paper
%%%%%%%%%%%%%%%%%%%%%%%%%%%%%%%%%%%%%%%%%%%%%%%%%%%%%%%%%

\section{
 Introduction}

 The  use of parton distribution functions (PDFs) $f(x)$ \cite{Feynman:1973xc}  
 is an  important tool to accumulate information about hadron structure. For many decades, 
 PDFs  were  the objects of intensive experimental studies, and now also of lattice QCD calculations  as well.
The gluon PDFs are the most difficult to investigate, both experimentally and on the lattice.  
 The ``classic'' gluon PDFs, unpolarized and polarized ones,
 are both related to twist-2 operators built from the gluon fields.
 Recently, X. Ji   proposed  \cite{Ji:2020baz} to consider  gluon PDFs  generated from
 the  twist-4 operators of $G^{+- } \ldots G^{+ -}$, etc.   type. 
  In Ref. \cite{Hatta:2020iin}, the twist-4 PDF $F(x)$ corresponding to the 
 $G_{\mu \nu} \ldots G^{\mu \nu}$  operators was introduced. 
 Its importance stems from the fact that the matrix element of the local operator 
 $\langle P | G_{\mu \nu} (0) G^{\mu \nu} (0)|P \rangle$ may be related to the gluon contribution into the proton mass. 

 An interesting question is whether $F(x)$ has a singular $\delta (x)$ part, sometimes  dubbed as a ``zero-mode''
 contribution. Such terms have been observed  \cite{Burkardt:2001iy}
 in one-loop perturbative QCD expressions for the 
 twist-3 quark PDFs.   The presence of such terms in twist-4 gluon PDFs 
  was suggested in Ref.   \cite{Ji:2020baz}. For  $F(x)$, this question   was   addressed in Ref. \cite{Hatta:2020iin}
through a one-loop calculation of the matrix element of the bilocal operator 
$G_{\mu \nu} (z) G^{\mu \nu} (0)$ between  virtual gluon states.

The calculation of Ref. \cite{Hatta:2020iin} was   performed in the light-cone gauge $(nA)=0$ 
and produced a $1/x$ term  in the evolution kernel.
It has emerged  from the $1/(kn)$ factor 
of the gluon propagator in the light-cone gauge. 
The authors of Ref. \cite{Hatta:2020iin} assumed that this $1/(kn)$ singularity 
should be supplied by the   Mandelstam-Leibbrandt  prescription \cite{Leibbrandt:1987qv}
which converts $1/x$ into $(1/x)_+$.
Formally,   the ``plus''  prescription for $1/x$ 
contains the  $\delta (x)$ term, and one may argue that this is an indication for 
  ``zero-mode'' terms in $F(x)$. 
However, the $x$-integral of  $(1/x)_+$  vanishes, while  the  genuine  ``zero-mode'' terms,
like those observed in Ref.  \cite{Burkardt:2001iy},  are expected to 
have a  pure $\delta (x)$ form   that  gives   a nonzero  contribution after integration.

A more essential  question is whether this $1/x$ term exists at all.
A worrisome fact is that  the calculation of Ref. \cite{Hatta:2020iin} was done using external gluons
with nonzero virtuality, which violates gauge invariance. 
A natural check would be  to calculate the same matrix element using another gauge,
which has not been done in Ref. \cite{Hatta:2020iin}. 
So, we did such a calculation using Feynman gauge and   obtained    a completely 
different result.  In particular, its  evolution kernel part 
 does not have the $1/x$ term found in the light-cone-gauge calculation,
but  contains a familiar/expected   $\sim [1/(1-x)]_+$ bremsstrahlung term absent in the result of Ref. \cite{Hatta:2020iin}. 

Our goal in the present paper is to revisit the issue of one-loop 
corrections for $F(x)$, and perform their calculation in a gauge-invariant way.

To  secure   gauge invariance, one needs  to  do the   calculations  
using  on-shell external gluons.
However,  the   tree-level  matrix element of the $G_{\mu \nu} (0) G^{\mu \nu} (z)$ operator 
 for such states  vanishes.  
 To avoid this problem, we have proposed to  take  a nonforward matrix element between 
 on-shell gluons with different lightlike momenta $p_1$ and $p_2=p_1+q$.  
 In other words, 
 we have 
considered 
the  generalized parton distribution (GPD)\footnote{GPDs have been  also used earlier \cite{Aslan:2018tff}  to investigate  ``zero modes'' 
in twist-3 quark distributions.  For twist-4 gluon $G \tilde G$-operators,  GPDs have been used in Ref. \cite{Hatta:2020ltd},
 however, for off-shell gluons still.}   $F(x,\xi,q^2)$ corresponding to the same bilocal operator $G_{\mu \nu} (0) G^{\mu \nu} (z)$ 
  as in  Eq. (\ref{tree}).  

We have found  that  the $\xi=0$ GPD $F(x,q^2)$ contains a $\delta (x)$ term.
However, it is accompanied by a $q^2$ factor and vanishes  in the  $q\to 0$ limit,
so that the ``forward'' PDF $F(x)$ does not have $\delta (x)$ terms at one loop. 
We have also obtained the one-loop evolution kernel for $F(x)$ and observed  
that it  is given by a  $\sim [1/(1-x)]_+$ bremsstrahlung  term and nothing else. 
No $1/x$ singularities in the evolution kernel have been detected. 
%Thus, the result of \mbox{Ref. \cite{Hatta:2020iin}}  is completely  misleading in this respect. 

The paper is organized in the following way. 
We start, in Section 2,  with the description of  results obtained 
using off-shell gluons. In Section 3, we introduce the GPD corresponding to a nonforward 
matrix element for on-shell gluons. In Section 4, we present our results for 
the main bulk of  diagrams, the ``real corrections'', and show that, for nonzero $q^2$,
they  contain the ``zero mode'' $\delta (x)$ term.  The formal origin 
of the $\delta (x)$ terms in one-loop integrals is investigated in Section 5. 
The structure of singularity at the $x=\pm 1$ endpoints is studied in Section 6.
We  separate this singularity into a  term that is regularized by the plus-prescription 
at $x=1$, and a term proportional to $\delta (1-x)$. 
  In Section 7,  we investigate  its structure and discuss 
 taking the $q^2\to 0$ limit.
Finally, in Section 8  we summarize the paper and formulate our conclusions.

\section{Regularization by external gluon virtuality}

 The  twist-4 PDF $F(x)$ may be defined \cite{Hatta:2020iin}   through a bilocal operator $G_{\mu \nu} (0) W[0,z] G^{\mu \nu} (z)$ on the light cone $z^2=0$
$$
F(x)=\frac{P^{+}}{2 M^{2} N_g} \int \frac{d z^{-}}{2 \pi} e^{i x P^{+} z^{-}}\left\langle P\left|G^a_{\mu \nu}(0) W[0, z^-] G_a^{\mu \nu}\left(z^{-}\right)\right| P\right\rangle \ , 
$$
where  $ W[0,z]$ is  the usual gauge link, $N_g=N_c^2-1$ is the number of gluons, and summation over the  hadron polarizations is implied.
The ``plus''-components are obtained by a scalar product with  a light-cone vector $n$, i.e.,  for an arbitrary vector $a$, one has  $n \cdot a=a^{+} .$ 

 In this definition of $F(x)$,   it is assumed  that there is no gluon propagator between the field points $-z^{-}/2$ and $z^{-}/2$,
 i.e. the $G$-fields enter  through a normal product. 
Also, it is implied that the $GG$ bilocal operator is in a $T$-product with the $S$-matrix,
i.e. one  deals with an  ``uncut'' PDF. This means that,  in perturbation theory,     all the diagram lines
correspond to usual propagators.  For  this reason,  
 the ``uncut'' PDFs    have the canonical support  $|x| \leq 1$  (see, e.g., Ref.  \cite{Radyushkin:1983wh}
 for an all-order proof).  Furthermore, 
the only foreseeable way to extract  these \mbox{twist-4} PDFs is through lattice simulations,  
and lattice calculations, of course,  involve just uncut propagators.

 The starting point of     perturbative calculations is a tree diagram corresponding to
  the matrix element of the twist-4 gluon operator
between gluon states with equal  \mbox{ momentum $p$.}
For  further generalizations, we take   different polarizations $\epsilon_{1}, \epsilon_{2}^*$ for these lines.
 At the tree-level, we have
\begin{align}
\frac{p^{+}}{N_g} &  \int \frac{d z^{-}}{2 \pi} e^{i x p^{+} z^{-}}\left\langle g\left(p, \epsilon_{2}^{*}\right)\left|G^a_{\mu \nu}(0) W[0, z] G_a^{\mu \nu}\left(z^{-}\right)\right| g\left(p, \epsilon_{1}\right)\right\rangle^{(0)} \nonumber \\
=& n \cdot p\left(p^{\mu} \epsilon_{1}^{\nu}-p^{\nu} \epsilon_{1}^{\mu}\right)\left(p^{\mu} \epsilon_{2}^{* \nu}-p^{\nu} \epsilon_{2}^{* \mu}\right) \delta(n \cdot p-x n \cdot p)+\{x\to -x\} 
%\delta^{ab}\delta^{ab} =2(N_c^2-1) 
\nn & 
= 2 \left(p^{2} \epsilon_{1} \cdot \epsilon_{2}^{*}-p \cdot \epsilon_{1}\,  p \cdot \epsilon_{2}^{*}\right) \delta(1-x)
+\{x\to -x\} \  . 
\label{tree}
\end{align}
One can see that,  for on-shell gluons,  i.e., when  $p^{2}=0$ and  
$p \cdot \epsilon_{1}=p \cdot \epsilon_{2}=0$,  the tree-level result  vanishes. 
%The one-loop result for the on-shell gluon states is zero as well. 
A non-vanishing result may be obtained if one takes off-shell gluons with  $p^{2} \neq 0$.
This has been done, in particular,  by the authors  of  Ref.  \cite{Hatta:2020iin}.   
 Such a choice has a certain  risk,  since  the one-loop  result may   be not  gauge invariant,
especially given the fact that the tree-level  expression  (\ref{tree}) is proportional to $p^2$, 
 the parameter characterizing the gauge invariance violation.

\begin{figure}[t]
%\centerline{\includegraphics[width=3in]{figaf}} %\vspace{-10mm}
\centerline{\includegraphics[width=5in]{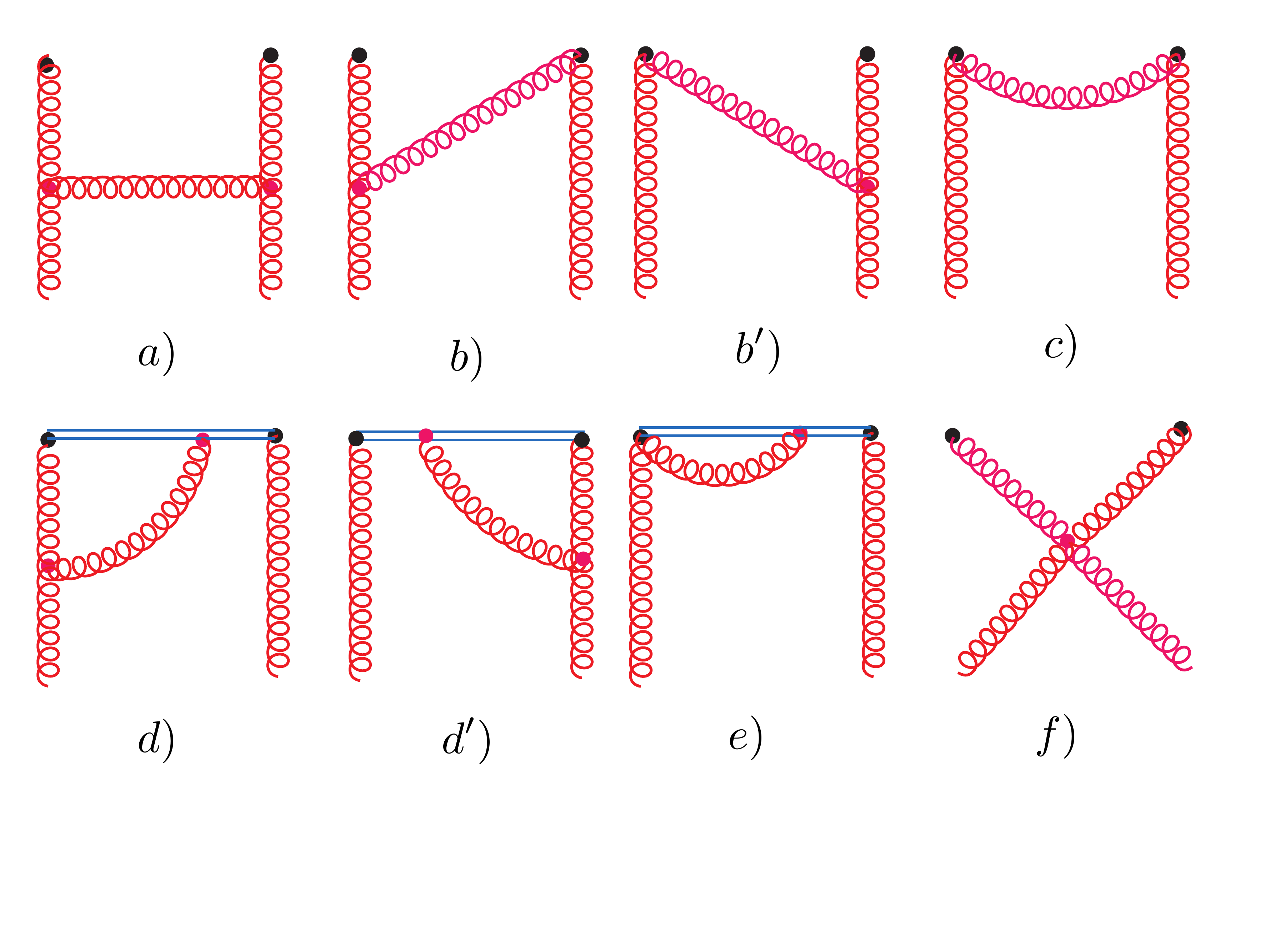}} %\vspace{-10mm}
\vspace{-15mm}
 \caption{ ``Real'' diagrams  (mirror diagram $e'$  is not shown). 
  \label{real}}
\end{figure}

For one-loop calculations, one should specify  the gauge.  In general, the gluon propagator is $-i D^{\mu \nu}(k) / k^{2}$, where
$D^{\mu \nu}(k)=g^{\mu \nu}$ in Feynman gauge  and 
$$
D^{\mu \nu}(k)=g^{\mu \nu}-\frac{k^{\mu} n^{\nu}+k^{\nu} n^{\mu}}{n \cdot k}
$$
 in the light-cone gauge.  Using the latter,   and the dimensional regularization  (DR) with $D=4-2 \epsilon$,  a one loop calculation 
  was performed  in  Ref.  \cite{Hatta:2020iin}. As we will discuss later, 
 the result has some peculiar features, so, for a check,   we redid this calculation  and   obtained 
\begin{align}
 F^{\rm LC}(x,p^2)\bigg|_{(a+b+b')} % \nonumber\\
&= - i C_A g^2 \frac{p^+}{-p^2}\int\frac{d k^{-}d^{D-2}\vec{k}_{\perp} }{(2\pi)^D}
\nn & \times 
\frac{(1-\epsilon)(1-x)[2k^- p^+(4-5x)+p^2 x(1+x)]+\vec{k}_{\perp}^2 [4(1-\epsilon)+x(6\epsilon-4)]}{(1-\epsilon)x(1-x)[\vec{k}_{\perp}^2-(1-x)(p^2-2k^- p^+)](\vec{k}_{\perp}^2-2k^- p^+ x)}  \nn & +(x \to -x) 
\end{align}
 for the sum of diagrams $a)$,   $b)$, $b')$  of Fig. \ref{real}. % ($b'$ being the mirror diagram of $b$).
This result coincides with Eq.~(45) of  Ref.  \cite{Hatta:2020iin}   if one expresses there 
 the denominators in the  light-cone components and substitutes  $k^+$ with $x p^+$.
  Taking the integrals and expanding in $\epsilon$ gives 
\begin{align}
F^{\rm LC}(x,p^2)\bigg|_{(a+b+b')}% \nonumber\\
& =- \frac{\alpha_s }{2\pi}C_A\bigg\{\bigg[\frac{1}{\epsilon}+\ln\frac{\mu^2}{-p^2x(1-x)}\bigg] \left (\frac{2}{x} -2+x\right )+x\bigg\} \   \theta(0< x <1 ) 
\nn & 
+(x \to -x)   \ .
\label{HZ}
\end{align}

The term $\sim 1/x$,  singular for $x=0$, comes from the light-cone-gauge denominator factor $1/(kn)$ which, 
surprizingly, is not cancelled  by  numerator factors  in this calculation. 
It is argued in Ref. \cite{Hatta:2020iin} that one should use here  
the  Mandelstam-Leibbrandt  prescription \cite{Leibbrandt:1987qv}
for the $1/(kn)$  factor,
which converts $1/x$ into $(1/x)_+$.
The ``plus''  prescription for $1/x$ 
formally contains a $\delta (x)$ term. 
A minor  comment here is that 
 the $x$-integral of $(1/x)_+$  vanishes, which is not what is usually expected from the ``zero-mode'' terms.
The latter  are believed to 
have a  pure $\delta (x)$ form for $x \sim 0$  that integrates to a nonzero  contribution.
Another strange feature of Eq. (\ref{HZ}) is the absence of a $1/(1-x)$ bremsstrahlung 
contribution, typical for PDFs in gauge theories like QCD.

To  check   if  it  is worth efforts  to    investigate   the puzzling structure of Eq. (\ref{HZ})  any further, 
we have calculated $F(x)$  
 in   Feynman gauge using,  again,   virtual external gluons.  We found that the result   is given by  a completely 
different analytic expression. In particular, for terms containing the evolution logarithm 
 $\ln (-\mu^2/p^2)$, we have obtained 
\begin{align}
F^{\rm F}(x)\bigg|_{(a+b+b' +d +d')}&=\frac{\alpha_s }{4\pi}C_A\bigg\{\bigg[\frac{1}{\epsilon}+\ln\frac{\mu^2}{-p^2 x(1-x)}\bigg]
\left ( \frac{2}{1-x}+1-x \right )-\frac{1+2x^2}{x}\bigg\}  
\nn & \times \theta(0< x <1 ) +(x \to -x)  \ .
\label{FG}
\end{align}
As one can observe, the evolution  term here does not have a $1/x$ singularity 
 for small $x$. However, 
it has   the usual $1/(1-x)$ soft radiation  singularity for $x=1$ absent in Eq.(\ref{HZ}). 
Still, in Eq. (\ref{FG}), we  see   a $1/x$ term in the non-evolution  part, with no evident prescription now 
how to regularize it. 

The drastic   difference between the Feynman- and  light-cone-gauge   results
%combined with the fact that both have been calculated in a  gauge-invariance-violating environment,
suggests that there are  good chances that neither  of them should be relied upon. 
A  natural suspect   for the origin of the discrepancy between the two  results and  of their peculiar features  is the 
 violation of  gauge invariance  stemming from   using virtual external gluons. 

\section{Regularization by nonforward kinematics for on-shell gluons}

To get a nonzero result at  the tree level, and preserve   gauge invariance at the one-loop level,  we have decided to 
take on-shell gluons, but impose a nonzero momentum transfer\footnote{The tree-level
 matrix element of the ``topological'' $F \ldots \tilde F$ operator,  considered in Ref.\cite{Hatta:2020ltd}, 
 vanishes even for $p^2 \neq 0$, so its  author took different gluon momenta $p_1$, $p_2$, still keeping both  $p_1^2$ 
 and $p_2^2$ nonzero.}  between them\footnote{
Another  way  is to 
calculate  corrections in the operator form, without  projections
on external states, like it was done  in Refs. \cite{Radyushkin:2017lvu,Balitsky:2019krf}  for
quark   and   gluon operators 
outside the   light cone.}.
This means that we
consider 
the  generalized parton distribution (GPD). 
The
non-forward matrix element   defining this GPD  is given by  
\begin{align}
F\left(x,  p_1, p_2; \epsilon_{1}, \epsilon_{2} \right)&=\frac{P^{+}}{N_g} \int \frac{d z^{-}}{2 \pi} e^{i x P^{+} z^{-}}
\nn & \times \left\langle g\left( p_2, \epsilon_{2}\right)\left|G^a_{\mu \nu}\left(-\frac{z^{-}}{2}\right) W\left[-\frac{z}{2}, \frac{z}{2}\right] G_a^{\mu \nu}\left(\frac{z^{-}}{2}\right)\right| g\left(p_1, \epsilon_{1}\right)\right\rangle \ ,
\end{align}
where  $p_1$ and $p_2$  are the on-shell ($p_1^2=p_2^2=0$) gluon momenta, 
 $P=(p_1+p_2)/2$ being  their average, and $q=p_2-p_1$ being their difference.
% The ``plus''-component of  a momentum  $k$ are obtained by their convolution $k\cdot n$ 
% with a lightlike vector $n$. 
The  skewness variable $\xi$ is defined  using    $p_1^+ = (1+\xi) P^{+}$ and $p_2^+ = (1-\xi) P^{+}$, so that 
   $\xi = -{q^{+}}/{2 P^{+}} .$ 
   %We also assume that $n \cdot p=(1+\xi) n \cdot P .$ So 
 %  Furthermore, we take $p_{1\perp} = q_{\perp}/2$, $p_{2\perp} = -q_{\perp}/2$.
 %  , hence,  $q_{\perp}^{2}=-q^{2}/(1-\xi^2)$.
The gluon polarizations satisfy the usual relation for on-shell gluons
$%\begin{align}
 \quad p_1 \cdot \epsilon_{1}=0, \quad p_2 \cdot \epsilon_{2}^{*}=0$ \  . 
 
%\end{align}
With  these definitions, we obtain  the following  tree-level result (recall  $P^+= n \cdot P$, etc.)
\begin{align}
F^{(0)}  \left(x,  p_1, p_2; \epsilon_{1}, \epsilon_{2} \right) & =n \cdot P\left(p_1^{\mu} \epsilon_{1}^{\nu}-p_1^{\beta} \epsilon_{1}^{\alpha}\right)\left( p_2^{\mu} \epsilon_{2}^{* \nu}- p_2^{\nu} \epsilon_{2}^{* \mu}\right)
\nn
& \times  [\delta(n \cdot P-x\,  n \cdot P) +\delta(n \cdot P+ x\, n\cdot P)]% \delta^{ab}\delta^{ab} 
\nn &
=2 \left(p_1 \cdot p_2 \,  \epsilon_{1} \cdot \epsilon_{2}^{*}-p_1 \cdot \epsilon_{2}^{*} \, p_2 \cdot  \epsilon_{1}\right)  [\delta(1-x) +\delta (1+x)]
\nn & 
=\left(-q^{2} \epsilon_{1} \cdot \epsilon_{2}^{*}+2 \, q \cdot \epsilon_{1} \,  q \cdot \epsilon_{2}^{*} \, \right) [\delta(1-x) +\delta (1+x)]
\nn & 
\equiv   \Pi (q,\epsilon_{1}, \epsilon_{2} ) [\delta(1-x) +\delta (1+x)] \  . 
\end{align}
We have used  here that $ p_2 \cdot \epsilon_{1}=  q \cdot \epsilon_{1}$ and $ p_1 \cdot \epsilon_{2}^*=  -q \cdot \epsilon_{2}^*$,  to make it  explicit 
that the structure $ \Pi (q,\epsilon_{1}, \epsilon_{2} )$  (and, hence, $F^{(0)}  \left(x,  p_1, p_2; \epsilon_{1}, \epsilon_{2} \right)$) vanishes in the forward limit  $q=0$. 

In  the nonforward case, we  parametrize $F\left(x,  p_1, p_2; \epsilon_{1}, \epsilon_{2} \right)$ 
by two invariant functions
\begin{align}
F  \left(x,  p_1, p_2; \epsilon_{1}, \epsilon_{2} \right) =  & \epsilon_{1}^{\ \mu}    \epsilon_{2}^{*\, \nu }\left [  \left(2 \, q_\mu
q_\nu -q^{2} g_{\mu \nu}  \right ) F_1 (x, \xi , q^2) + g_{\mu \nu}   F_2 (x, \xi , q^2)   \right ] 
\nn & 
\equiv   \Pi (q,\epsilon_{1}, \epsilon_{2} )  F_1 (x, \xi , q^2) +  \epsilon_1\cdot \epsilon_2^*  F_2 (x, \xi , q^2) \ .
\end{align}
The Lorentz structure of the first term coincides with that observed at the tree level. 
It is produced  by a traceless 2-dimensional tensor when  $\alpha, \beta$ are chosen to be in the transverse plane. 
The second term  corresponds to the trace in  such  2-dimensional indices. 
Thus, at the tree level, we have $ F_1^{(0)} (x, \xi , q^2)  =  \delta(1-x) +\delta (1+x) $ and $ F_2^{(0)} (x, \xi , q^2) =0$.

\section{Total result for ``real corrections''}

We have calculated the  diagrams $1a) - 1f)$  (sometimes called   ``real corrections'') both 
 in Feynman and light-cone gauges (the details of these rather lengthy calculations will be presented in a separate paper
 \cite{Zhao}). In both gauges,   we  have obtained  the same total expression.
 
 %(the details of these rather involved calculations will be presented elsewhere). 
 Let us discuss  first the results for  $ F_{1,2}^{(1)} (x, \xi , q^2)$ away from the potentially singular endpoints  $x = \pm 1$. 
  We start with the function $ F_1\left(x, \xi, q^{2}\right)$ that was nonzero at the tree level. 
 At  one loop, it  has the following form 
 \begin{align} 
F_1^{(1)}  \left(x, \xi, q^{2}\right) & \left.    \right |_{x \neq \pm 1}= \frac{\alpha_{s}}{\pi} C_{A}
 \frac1{1-x}   \left\{
%\left[
%\frac{2 q \cdot \epsilon_{2}^{*} q \cdot \epsilon_{1}-q^{2} \epsilon_{1} \cdot \epsilon_{2}^{*}}
\left(\frac{1}{\epsilon_{\rm UV}}+\ln \frac{\mu_{\rm UV}^{2}\left(1-\xi^{2}\right)}{-q^{2}(1-x)^{2}}\right)
%+q^{2} \epsilon_{1} \cdot \epsilon_{2}^{*} \frac{1-x}{1-\xi^{2}}
%\right]
 \theta(\xi < x < 1)\right.\nn 
 & \left.+
 %\left[
% {2 q \cdot \epsilon_{2}^{*} q \cdot \epsilon_{1}-q^{2} \epsilon_{1} \cdot \epsilon_{2}^{*}}
 %{2(1-x)}
 \frac12 
 \left(\frac{1}{\epsilon_{\rm UV}}+\ln \frac{\mu_{\rm UV}^{2}(\xi-x)(1+\xi)^{2}}{-q^{2}(1-x)^{2}(x+\xi)}\right)
 %-q^{2} \epsilon_{1} \cdot \epsilon_{2}^{*} \frac{1-\xi}{4 \xi(1+\xi)}
% \right]
  \theta(-\xi < x < \xi)\right\} +\{x \to -x\} \  .
  \label{F1real}
 \end{align}
 As expected, it   contains  the evolution contribution   revealed by the 
$1/\epsilon_{\rm UV}$  pole and  the
  logarithmic  $\ln (-\mu_{\rm UV}^2/q^2)$ dependence on the 
UV  regulator scale $\mu_{\rm UV}$. 

One can see that 
 the momentum transfer $q^2$ serves here  as an IR cut-off,
 a subtlety that  we will briefly     address now and    in more detail later on.
The point is that, at intermediate stages of the calculations, we also had the integrals diverging both on UV and and IR sides and 
resulting in the 
poles $1/\epsilon_{\rm UV} - 1/\epsilon_{\rm IR}$ and 
 logarithms of $\ln (\mu^2_{\rm UV}/\mu^2_{\rm IR})$ type.
However, all the  poles $ 1/\epsilon_{\rm IR}$ and   the dependence  on $\mu_{\rm IR}$ cancel  in the final result.
A similar observation was made in the studies of the quark GPDs \cite{Liu:2019urm} (see also \cite{Ji:2015qla}).

Since we are interested in the  ``forward'' PDFs, we take the $\xi \to 0$  limit, to  get 
\begin{align}
\left. F_1^{(1)} \left(x,  q^{2}\right) \right |_{x \neq \pm 1} =
 \frac{\alpha_{s}}{\pi} C_{A}
 & \left\{     \frac{\theta(0\leq x<1) }{1-x} \left(\frac{1}{\epsilon_{\mathrm{UV}}}
 +\ln \frac{\mu_{\mathrm{UV}}^{2}}{-q^{2}(1-x) ^2} \right) 
% \right. \nn &\left.
 % +q^{2} % \epsilon_{1} \cdot \epsilon_{2}^{*}
%   \left  ((1-x)\theta (0\leq x\leq 1)-\frac12 \delta( x) \right)
    \right\} 
    \nn & 
+\{ x \to -x \} \ .
\label{xi01}
\end{align}
 Here and in what follows, 
 we denote $F \left(x, \xi =0,  q^{2}\right) = F  \left(x,  q^{2}\right)$ 
 (similarly, we will denote later $F \left(x,   q^{2}=0 \right) = F  \left(x \right)$).
 
 Recall that   the function $ F_2 \left(x, \xi, q^{2}\right)$  vanishes at the tree level. Thus, 
one would expect
%   before  any calculations,    
 that  it should   not contain  evolution logarithms at one loop. 
This expectation is supported by the actual  one-loop  result 
 \begin{align} 
  F_2^{(1)} \left(x, \xi, q^{2}\right)= \frac{\alpha_{s}}{\pi} C_{A}
&q^{2}   \left\{
%\left[
%\frac{2 q \cdot \epsilon_{2}^{*} q \cdot \epsilon_{1}-q^{2} \epsilon_{1} \cdot \epsilon_{2}^{*}}
\frac{1-x}{1-\xi^{2}}
%\right]
 \theta(\xi < x \leq 1)
 %\right.\\  & \left.
 %\left[
% {2 q \cdot \epsilon_{2}^{*} q \cdot \epsilon_{1}-q^{2} \epsilon_{1} \cdot \epsilon_{2}^{*}}
 %{2(1-x)}
 -%\frac12 
\frac{1-\xi}{4 \xi(1+\xi)}
% \right]
  \theta(-\xi <x < \xi)\right\} 
  \nn &  +\{x \to -x\} \  .
    \label{F2real}
 \end{align}
 We have dropped here  the $x \neq \pm 1$ restriction because $F_2^{(1)} \left(x, \xi, q^{2}\right)$ is not singular 
 (in fact, vanishes) 
 for $x=\pm 1$. 
 Taking   the  $\xi \to 0$  limit of this expression
is not straightforward, because 
  it has a singular  $\sim 1/\xi$ behavior 
in the central  region $-\xi < x <  \xi$.   
As one can notice,  $\theta(-\xi < x < \xi)/2\xi$ converts into  a ``zero mode'' $ \delta (x)$  term  in the $\xi \to 0$ 
limit\footnote{Similar structure is present in the  twist-3 quark GPDs studied  in Ref. \cite{Aslan:2018tff}.}.
Using this observation and taking the $\xi \to 0$  limit, we get 
\begin{align}
F_2^{(1)} \left(x,  q^{2}\right)   =
& \frac{\alpha_{s}}{\pi} C_{A} q^{2}  \left\{   (1-x)\theta (0\leq x\leq 1)-\frac12 \delta( x) \right\} 
+\{ x \to -x \} \ . 
\label{xi02}
\end{align}

Clearly,  the kinematics with $\xi=0$ may    still  be non-forward, as far   as  $q^2 \neq 0$.
% It   corresponds to a situation when $q$ is purely transverse, \mbox{$q=\{ 0,0, q_\perp \}$. } 
 Thus, one can try  to  calculate   $F_{1,2} (x, q^2) $  imposing  the $\xi=0$  condition   from the start. 
 We did such a calculation,  
 both in the light-cone and Feynman 
gauges. At the end,  we have   obtained identical total results coinciding with Eqs. (\ref{xi01}),  (\ref{xi02}).   
%We want to note here that   
 In fact,   such an  outcome is not completely  trivial, since we have observed that 
the results obtained from the 
$\xi\to 0$ limit of the $\xi \neq 0$ calculation sometimes  differ on the diagram-by-diagram level 
  from those obtained when
$\xi$  equals  zero from the beginning.  Only the combined results coincide.

Thus,    the only singularity for $x=0$ in the one-loop results (\ref{xi01}),   (\ref{xi02})
 is the $\delta(x) $ term in $F_2 \left(x,  q^{2}\right) $. 
All the other terms are not singular for $x=0$. Note, however, that  the function $F_2^{(1)} \left(x,  q^{2}\right) $  
vanishes in the forward limit, when $q=0$. Thus, our    perturbative  one-loop calculation does not indicate  presence 
of the ``zero-mode'' terms in the forward PDF $F(x)$.
Still, one may say that  the ``zero-mode''  $\delta (x)$  term is   present for the $\xi=0$ GPD 
 in the $F_2^{(1)} \left(x,  q^{2}\right) $  term  when $q^2 \neq 0$.
As a side remark, we note also   that  the $x$-integral of  $F_2^{(1)} \left(x,  q^{2}\right) $ vanishes.

%The $\delta (x)$  contribution comes  from momentum integrals of the ``$D_1 D_3$'' type, 
%where $D_1 = k^2$, $D_2 = (p-k)^2$, $D_3=(k+q)^2$   are the denominators of 
%the three types of propagators  that one encounters in this one-loop calculation. 
%The simplest of ``$D_1 D_3$''  integrals is given by
%\begin{align}
%S(1,0,1) = \int \frac{d^D k} {(2\pi)^D} \frac{\delta (x- k\cdot n/ P\cdot n)} {k^2 (k+q)^2}  = \delta (x) 
%\end{align}

\section{``Zero modes''  in one-loop integrals}

A  frequent argument for ``zero modes'', i.e.,  $\delta (x)$ terms  in parton  distributions,  is based on the results  of 
 perturbative one-loop calculations \cite{Burkardt:2001iy,Aslan:2018tff,Bhattacharya:2020jfj}.  
 It is interesting to trace the mechanism that leads to the $\delta (x)$ terms in the contributions of particular diagrams.  
 
 In our case, the ``zero mode'' terms  in  $F(x, q^2)$ have been    produced, in particular,  by the box diagram $1a)$. 
Consider a   generic one-loop   integral  corresponding to a  box    diagram in  GPD kinematics,  given by
\begin{align} 
I_{n_1 n_2 n_3}(x, \xi, q^2) \sim \int d^D k \frac{P(k, p_1,p_2)}{(p_1-k)^{2n_1} k^{2n_2} (p_2-k)^{2n_3}} \delta  (1-x- k^+ / P^+) \ , 
\label{bastria}
 \end{align}
where the function $P(k, p_1,p_2) $ comes from   numerator factors. ``Zero modes'' appear when  $P(k, p_1,p_2)$ is proportional to 
 $k^{2n_2}$,  which   cancels the middle propagator factor $1/k^{2n_2}$.  The resulting integral does not 
 depend on the virtualities of external   momenta $p_1^2, p_2^2$. Only the dependence on  $q^2=(p_2-p_1)^2$  (and $\xi$)  remains.
 Take  the simplest case  
 when \mbox{$n_1=n_3=1$}, $n_2=0$ and $P(k, p_1,p_2)=1$, which we denote as $S_{101}$, 
\begin{align} 
S_{101}(x, \xi, q^2) \sim \int d^D k \frac{  \delta  (1-x- k^+ / P^+)}{(p_1-k)^2  (p_2-k)^2}\ .
\label{Fxal0}
 \end{align}
 Writing the denominator factors in the Schwinger $\alpha$-representation, we have,   in the $\overline{\rm MS}$ scheme, 
 \begin{align} 
S_{101}(x, \xi, q^2) =   (\mu^2 e^{\gamma_E})^{\epsilon}   \int_0^\infty \frac{d\alpha_1 d\alpha_3}{( \alpha_1 +\alpha_3)^{2-\epsilon}}  \delta  \left (x - \frac{ \xi (\alpha_3 - \alpha_1 )  }{\alpha_1 +\alpha_3} \right ) e^{q^2 \alpha_1 \alpha_3/(\alpha_1 +\alpha_3)} \ \ .
\label{Fxal}
 \end{align}
Rescaling $\lambda= (\alpha_1+\alpha_3) \ , \  \beta = \alpha_1/(\alpha_1+\alpha_3) $ gives 
\begin{align} 
S_{101}(x,\xi,q^2  ) =  (\mu^2 e^{\gamma_E})^{\epsilon}  \int_0^\infty \frac{d \lambda}{\lambda^{1-\epsilon}}   \int_0^1 d \beta \,  \delta  \left (x -  \xi (1-2 \beta)   \right ) e^{q^2 \lambda \beta (1-\beta)}   \ .
 \end{align} 
Since $|1-2 \beta|\leq 1$, this contribution vanishes for $|x|>\xi$, so it  exists 
in the middle region  $|x|<\xi$  only (similar results have been obtained in calculations of twist-3 quark GPDs \cite{Aslan:2018tff}), 
\begin{align} 
S_{101}(x,\xi,q^2  )= \frac{ \theta(|x|<\xi) } 
{2 \xi}  (\mu^2 e^{\gamma_E})^{\epsilon}   \int_0^\infty \frac{d \lambda}{\lambda^{1-\epsilon}}    e^{\lambda q^2 (1-x^2/\xi^2)/4} \to  \frac{ \theta(|x|<\xi) } {2 \xi} \ln (-q^2/\mu^2)  + \ldots  \ \ .
\label{s101f}
 \end{align}
 Taking the $\xi \to 0$  limit of $\theta (|x|<\xi)/2 \xi$ gives $\delta (x)$, so $I_{101}(x,q^2) \sim \delta (x) \ln (-q^2/\mu^2)  + \ldots $ . 
 This result may  also  be obtained by simply   substituting  $\xi =0$ in Eq. (\ref{Fxal}).
  Note that the non-regularized  $\lambda$-integral diverges  in the  region of   small $\lambda$, 
 i.e. in the UV region. Hence the  $\mu$-parameter here has the meaning of $\mu_{\rm UV}$,
 with $-q^2$ serving as the IR cut-off in the large-$\lambda$ region.
 
 The ``zero mode'' terms  in  $F(x, q^2)$ may be    produced also  by  the four-gluon-vertex diagram $1f)$. 
  In this case, the $1/k^2$ propagator is absent
 from the beginning. 
 In the light-cone  gauge, we found that 
 the ``zero mode'' terms  $\delta (x)$   accompanied by $\ln (-q^2/\mu^2)$  factors  are present in $F_1(x, q^2)$
  for  diagrams $1a)$ and $1f)$, but they  cancel each other. 
 The surviving  ``zero mode'' term in   $F_2(x, q^2)$ has  resulted  from the  diagram $1f)$.
 It comes   from the   integral  (\ref{s101f}) 
 multiplied  by an extra $\epsilon$ factor, which cancels the $1/\epsilon$
 singularity produced by the integral over $\lambda$. 
In addition,  there is a numerator  $q^2$  factor  accompanying this integral, which results in the  $\sim q^2$ overall factor
for  the ``zero mode'' contribution to $F_2(x, q^2)$.

%\newpage 
\section{
Structure of    singularity at $x=1$ }

Another domain,  where one may expect singularities in QCD,  is the soft-gluon region $x\sim 1$ (or $x\sim -1$), in which 
 the external momentum is wholly carried by the active parton. 
Since  the function $F(x,q^2)$ is even in $x$, it is sufficient to discuss its $x\geq 0$ part.

In our case, 
the function $F_1^{(1)} \left(x,  q^{2}\right) $ in Eq. (\ref{xi01})   has a $1/(1-x)$ singularity in its evolution kernel.
In the lightcone gauge, it comes solely from Fig. $1a)$, which produces the $1/(1-x)$ singularity  in a ``bare'' form, 
without   a plus-prescription for it, while   the  accompanying  $\delta (1-x)$  terms come from 
self-energy diagrams.

In Feynman gauge,  the diagram $1a)$  also produces $\sim 1/(1-x)$ terms, but they come, in addition,   from  the  diagrams $1d)$  and $1d'$) as well.
These diagrams    contain a gluon insertion into the gauge-link line. They 
produce equal contributions which, in fact,  have the  plus-prescription form for the singularity at $x=1$. Their sum is given by 
%{\color{red}
\begin{align}
 F^F_{(1d+1d')} \left(x, q, \epsilon_1, \epsilon_2\right)= &\frac{\alpha_s}{2\pi}C_A
 \left (\frac{1}{\epsilon_{\mathrm{UV}}}-\frac{1}{\epsilon_{\mathrm{IR}}}+\ln\frac{\mu_{\mathrm{UV}}^2}{\mu_{\mathrm{IR}}^2}\right )
  \Pi (q,\epsilon_{1}, \epsilon_{2} ) 
 \left [\theta(0\leq x \leq 1)\frac{x+x^2}{1-x} \right ]_+ \nn & +\{x \to -x\} \ .
 \label{Fdd'} 
\end{align}
%}
Note that the momentum integrals in  these diagrams do not depend on the momentum transfer $q$,
 and diverge both at 
the UV and IR ends of integration, the fact reflected in the overall factor containing the poles. 

The   box diagram $1a)$ in Feynman gauge 
\begin{align}
F^F_{(1a)}(x, q, \epsilon_1, \epsilon_2)=&\frac{\alpha_s}{2\pi}C_A \bigg\{ \Pi (q,\epsilon_{1}, \epsilon_{2} ) 
\bigg[
 (2+x)\left (\frac{1}{\epsilon_{\mathrm{UV}}}-\frac{1}{\epsilon_{\mathrm{IR}}}+\ln\frac{\mu_{\mathrm{UV}}^2}{\mu_{\mathrm{IR}}^2} \right )
\nn &
 -
\frac{\Gamma(-\epsilon_{\mathrm{IR}})^2\Gamma(1+\epsilon_{\mathrm{IR}})}{\Gamma(-2\epsilon_{\mathrm{IR}})}
\left (\frac{\mu_{\mathrm{IR}}^2 e^{\gamma_E}}{-q^2} \right )^{\epsilon_{\mathrm{IR}}}
(1-x)^{-1-2\epsilon_{\mathrm{IR}}} 
\bigg ]
%\nn   &
%\theta(0 \leq x\leq1)
 \nn
& + 2q^2   \epsilon_1\cdot \epsilon_2^*
(1-x) \bigg\}  \theta(0<x \leq 1)  +\{x \to -x\} + \delta (x) \ {\rm terms} \  ,
\label{1a}
\end{align}
in its $ \Pi (q,\epsilon_{1}, \epsilon_{2} ) $ part, 
has  two types of contributions for nonzero $x$.
The first of them has the structure  similar to those of diagrams $1d)$  and $1d')$ ,  i.e.,    has both UV and IR poles. 

Contributions of  the second type  are UV finite and 
have  IR poles only.
%, which   correspond to collinear singularities due to zero virtualities  of  external gluons. 
They are contained in the  term with the gamma-functions in  the second line of Eq.(\ref{1a}). 
This term  corresponds 
to the basic integral (\ref{bastria}) with   {$n_1=n_2=n_3=1$ and $P(k,p_1,p_2)=-q^2$.}
We denote it as $(-q^2 )S_{111} (x, q^2)$. 
 This integral is UV finite, but has collinear divergences due to vanishing  virtualities $p_1^2=0$ and $p_2^2=0$  of external lines.  
To regularize these divergences,  we  have applied    dimensional regularization in the $\overline{\rm MS}$ scheme.
Using Schwinger representation, the regularized $S_{111} (x, q^2)$ function 
may be written as
 \begin{align} 
S_{111}(x,  q^2; \mu_{\rm IR}^2,\epsilon_{\rm IR})& 
= (\mu_{\rm IR}^2 e^{\gamma_E})^{\epsilon_{\rm IR}}  
% \frac1{\Gamma(1+\epsilon_{\rm IR}) } 
  \int_0^\infty {d\lambda} \lambda^{\epsilon_{\rm IR}} \int_0^1 d\beta_1 d\beta_2 d\beta_3\,  \delta  (1-\sum_{i=1}^3  \beta_i  )  \, \delta  \left (x - \beta_2 
 \right ) e^{\lambda q^2 \beta_1 \beta_3}  \ . 
 \label{s111lam}
 \end{align}
As  witnessed by  Eq. (\ref{1a}), this  integral   has singularities for $x=1$. 
 We would like to represent them  as a sum of a  term with the  plus-prescription at $x=1$ and a term proportional to $\delta (1-x)$.
 To this end,  we perform such a decomposition for 
 the   $(1-x)^{-1-2\epsilon_{\mathrm{IR}}}$ term
  \begin{align} 
 \frac{\theta (0\leq x\leq 1)}{ (1-x)^{1+2\epsilon_{\mathrm{IR}}}} = \left [  \frac{\theta (0\leq x\leq 1)}{ (1-x)^{1+2\epsilon_{\mathrm{IR}}}} \right ]_+
% {\color{red}-} 
 -\frac{1}{2\epsilon_{\mathrm{IR}}}
 \delta (1-x) \ . 
\label{s111do}
 \end{align}
 As a result, we have 
%{\color{red}
    \begin{align} 
(-q^2) S_{111}(x,  q^2; & \mu^2,\epsilon_{\rm IR})  
=
\left [  \frac{\theta (0\leq x\leq 1)}{1-x}\left ( \frac{2}{\epsilon_{\rm IR}
   }  -  2\ln \frac{-q^2(1-x)^2}{\mu_{\rm IR}^2} \right )  \right  ]_+  
\nn &
+
 \delta (1-x) 
\bigg( - \frac{1}{\epsilon_{\rm IR}^2}+ \frac{1}{\epsilon_{\rm IR}}\ln\frac{-q^2}{\mu_{\mathrm{IR}}^2}
	- \frac{1}{2}\ln^2\frac{-q^2}{\mu_{\rm IR}^2}+\frac{\pi^2}{12}\bigg) + {\cal O}(\epsilon_{\rm IR}) \ . 
	\label{s111do2}
 \end{align}
%}
Using the ``$[...]_+ + \delta(1-x)$'' form for the $~(2+x)$ term
in the first line of Eq.  (\ref{1a}),  and combining  contributions of 1$a$), 1$d$) and 1$d'$) diagrams
(diagrams $1b), 1b')$ and $1c)$ have vanishing contributions)  gives 
  \begin{align}
F^F_{(1a+1d+1d')}(x, q, \epsilon_1, \epsilon_2) = \frac{\alpha_s}{\pi}C_A 
&\bigg\{ 
  \Pi (q,\epsilon_{1}, \epsilon_{2} ) 
\bigg[\frac{\theta(0\leq  x\leq 1)}{1-x} 
\left (\frac{1}{\epsilon_{\mathrm{IR}}} + \ln \frac{\mu_{\mathrm{IR}}^2 }{-q^2(1-x)^2}   \right ) \nn
 &+\frac{\theta(0\leq  x\leq 1)}{1-x} 
\left (\frac{1}{\epsilon_{\mathrm{UV}}}-\frac{1}{\epsilon_{\mathrm{IR}}}+\ln\frac{\mu_{\mathrm{UV}}^2}{\mu_{\mathrm{IR}}^2} \right )
\bigg ]_+     
\nn   &
 +q^2 \epsilon_1\cdot \epsilon_2^*(1-x)\theta (0\leq x\leq 1)\bigg\}
\nn &
+ \delta (1-x)   \ {\rm term} +\{x \to -x\}  +  \delta (x) \ {\rm terms}  \ .
\label{1add}
\end{align}
We see  that the IR terms $1/\epsilon_{\mathrm{IR}}$  and $\ln \mu_{\mathrm{IR}}^2$    cancel 
and we get 
  \begin{align}
F^F_{(1a+1d+1d')}(x, q, \epsilon_1, \epsilon_2) =&\frac{\alpha_s}{\pi}C_A \bigg\{ 
  \Pi (q,\epsilon_{1}, \epsilon_{2} ) 
\bigg [\frac{\theta(0\leq  x\leq 1)}{1-x} \bigg(
\frac{1}{\epsilon_{\mathrm{UV}}} + \ln \frac{\mu_{\mathrm{UV}}^2 }{-q^2(1-x)^2}  
\bigg )\bigg ]_+
\nn   &
 +q^2 \epsilon_1\cdot \epsilon_2^* (1-x)\theta (0\leq x\leq 1)  \bigg\}
 \nn &
+ \delta (1-x)  \ \ {\rm term} +\{x \to -x\}  +  \delta (x) \ {\rm terms} \ . 
\label{pplusd}
\end{align}
A similar cancellation of  $1/\epsilon_{\mathrm{IR}}$  and $\ln \mu_{\mathrm{IR}}^2$ terms
was observed in the calculation of the matching conditions for quark quasi-GPDs in Ref. 
\cite{Liu:2019urm}.

Adding the Feynman-gauge-contribution of the diagram  $1f)$  (which is proportional to
$\delta (x)$)  gives the final result displayed  in Eqs. (\ref{xi01}),   (\ref{xi02}).

\section{Structure of $\delta (1-x) $ terms and $q^2 \to 0$ limit}

The sum (\ref{pplusd})  of all  ``real diagrams'' $1a)$ -- $1f)$ is given by a part having the plus-prescription form at $x=1$  and 
the  $\sim \delta (1-x)$ contribution  given by  the sum of the 
``ultraviolet'' term 
  \begin{align} 
    &\frac{\alpha_s}{2\pi}C_A \delta (1-x) 
\frac52 \left (\frac{1}{\epsilon_{\mathrm{UV}}}-\frac{1}{\epsilon_{\mathrm{IR}}}+\ln\frac{\mu_{\mathrm{UV}}^2}{\mu_{\mathrm{IR}}^2} \right )
	   \Pi (q,\epsilon_{1}, \epsilon_{2} )  \  
	\label{UV}
 \end{align}
 resulting from the $x$-integral of the $\sim (2+x)$ term in the first line of Eq. (\ref{1a}),
and the 
``Sudakov''  term
%{\color{red}
    \begin{align} 
    &\frac{\alpha_s}{2\pi}C_A \delta (1-x) 
\bigg(- \frac{1}{\epsilon_{\rm IR}^2}+ \frac{1}{\epsilon_{\rm IR}}\ln\frac{-q^2}{\mu_{\mathrm{IR}}^2}
	- \frac{1}{2}\ln^2\frac{-q^2}{\mu_{\rm IR}^2}+\frac{\pi^2}{12}\bigg)	
	 \Pi (q,\epsilon_{1}, \epsilon_{2} ) \  . 
	\label{Sud}
 \end{align}
 given by the second line of Eq.(\ref{s111do2}).
% Let us discuss them separately.
 %  \subsection{UV divergent terms}
   \begin{figure}[t]
%\centerline{\includegraphics[width=3in]{figaf}} 
\vspace{-4mm}
\centerline{\includegraphics[width=6in]{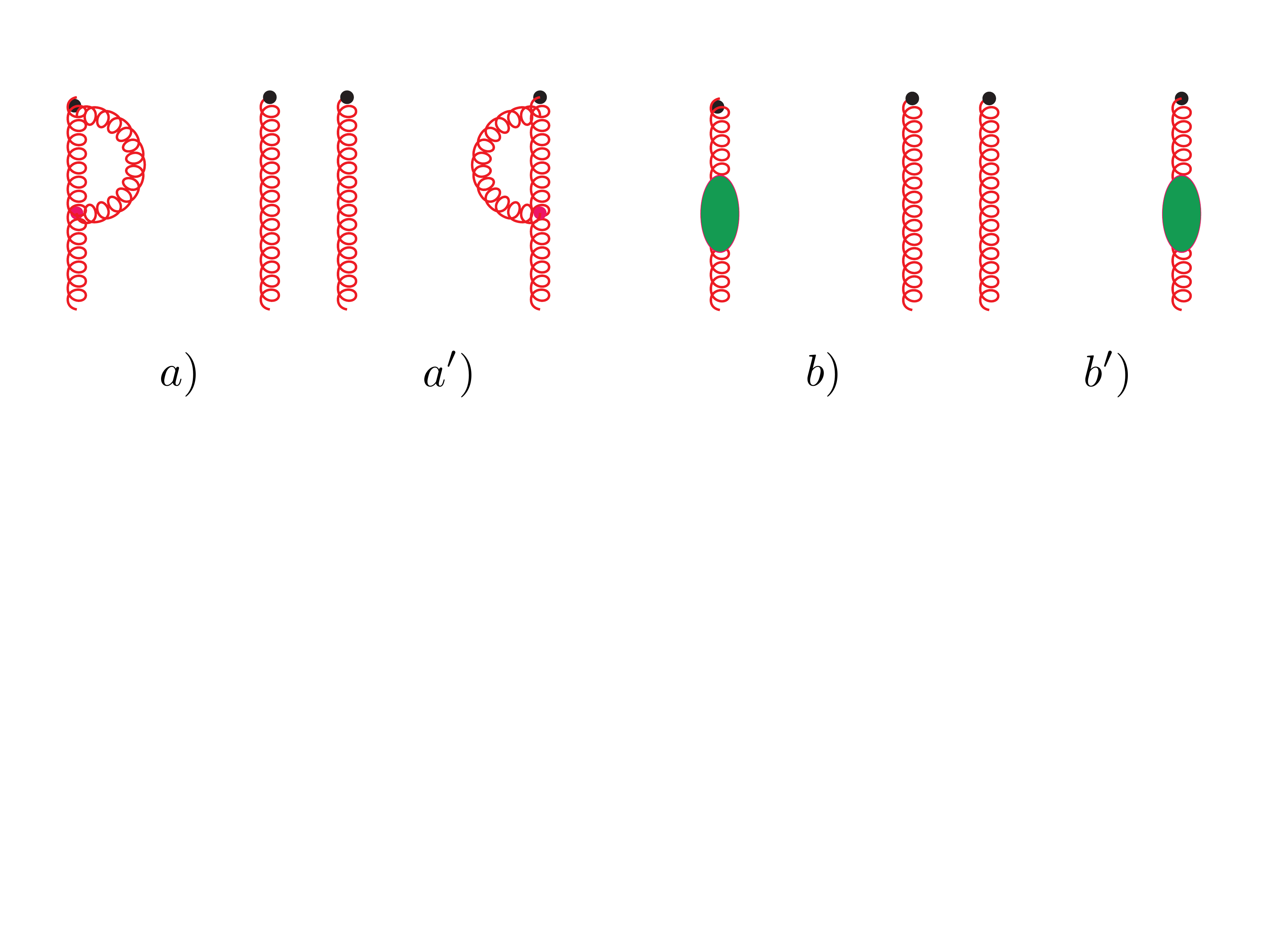}} %\vspace{-10mm}
\vspace{-55mm}
 \caption{Self-energy-type diagrams  \label{fig:self}}
\end{figure} 

UV-divergent terms   similar  to those in Eq. (\ref{UV}) are also present in the gluon self-energy-type diagrams 
    shown in Fig.~\ref{fig:self}.
  %In Feynman gauge, t
  The diagrams $2a)$ and $2a')$  together give
    \begin{align}
	F_{(2a+2a')}(x, q, \epsilon_1, \epsilon_2)=&-\delta(1-x)\frac{\alpha_s}{\pi} C_A \frac34 \left(\frac{1}{\epsilon_{\mathrm{UV}}}-\frac{1}{\epsilon_{\mathrm{IR}}}+\ln\frac{\mu_{\mathrm{UV}}^2}{\mu_{\mathrm{IR}}^2} \right) \Pi (q,\epsilon_{1}, \epsilon_{2} ) 
\nn & 	+\{x \to -x\} \ .
	\label{selins} 
	\end{align}
The gluon self-energy diagrams $2b), 2b')$  give
\begin{align}
	F_{(2b+2b')} (x, q, \epsilon_1, \epsilon_2 )& = \delta(1-x) \frac{\alpha_s}{\pi}
	\left (\frac{5}{12}C_A-\frac13 T_F n_f  \right )  \left(\frac{1}{\epsilon_{\mathrm{UV}}}-\frac{1}{\epsilon_{\mathrm{IR}}}+\ln\frac{\mu_{\mathrm{UV}}^2}{\mu_{\mathrm{IR}}^2}\right) \Pi (q,\epsilon_{1}, \epsilon_{2} ) \nn & 	 +\{x \to -x\} \ .
	\label{self}
\end{align} 
Combining    Eqs.  (\ref{UV}),   (\ref{selins})
and  (\ref{self})  we obtain 
\begin{align}
&	\frac{\alpha_s}{2\pi}\frac52 C_A-\frac{\alpha_s}{\pi}\frac34 C_A+\frac{\alpha_s}{\pi} \left(\frac{5}{12}C_A-\frac13 T_F n_f \right)  %\frac{1}{\epsilon_{\mathrm{UV}}}
 %\Pi (q,\epsilon_{1}, \epsilon_{2} )
% \nonumber\\
=\frac{\alpha_s}{4\pi} \left(\frac{11}{3} C_A-\frac43 T_F n_f\right)
%\frac{1}{\epsilon_{\mathrm{UV}}}  \Pi (q,\epsilon_{1}, \epsilon_{2} ) 
 % \nonumber\\
=\frac{\alpha_s}{4\pi}\beta_0  %\frac{1}{\epsilon_{\mathrm{UV}}} \Pi (q,\epsilon_{1}, \epsilon_{2} )\ .
\label{beta0}
\end{align}
as the coefficient accompanying $ \left({1}/{\epsilon_{\mathrm{UV}}}-{1}/{\epsilon_{\mathrm{IR}}}
+\ln ({\mu_{\mathrm{UV}}^2}/{\mu_{\mathrm{IR}}^2})
\right)$. 
 This result is in agreement with the fact that the combination $G^2 \beta (\alpha_s) /\alpha_s$ 
 is related to  the trace anomaly \cite{Chanowitz:1972vd,Crewther:1972kn,Collins:1976yq}, and 
 $g^2G_{\mu \nu} (0)G^{\mu \nu}(0)$  does not depend on the UV scale  $\mu_{\rm UV}$ at one loop \cite{Kluberg-Stern:1975ebk,Nielsen:1977sy}.  Hence, 
 the anomalous dimension of $G^2$   at one loop 
should be  opposite to that of $g^2$, i.e., proportional to  $\beta_0$ (see, e.g., Ref.  \cite{Tarrach:1981bi} ).

%\subsection{``Sudakov''  terms}

The ``Sudakov''  term  of  Eq. (\ref{Sud})  is UV finite. It does not contain the  UV parameter  
$\mu_{\mathrm{UV}}$, and thus it does not affect 
the relation between the functions $F_1^{(1)} (x,q^2,\mu_{\mathrm{UV}}^2)$ at different 
evolution scales $\mu_{\mathrm{UV}}$, which we will denote simply as $\mu_i$. Namely, for $x\geq 0$ we have 
   \begin{align}
F_{1}^{(1)} (x, q^2,\mu_1^2) =&F_{1}^{(1)} (x, q^2,\mu_2^2) +
\frac{\alpha_s}{\pi} \,
\left \{
C_A \bigg [\frac{\theta(0\leq  x\leq 1)}{1-x}\bigg ]_+ 
+ \frac{\beta_0}{4} \delta(1-x)
\right \} \ln \frac{\mu_1^2 }{\mu_2^2} 
%\nn   &
% \bigg\}
% \nn &
\label{evo1}
\end{align}
and similarly for  $x\leq 0$.
As indicated earlier,   the $F_2^{(1)} \left(x,  q^{2}\right)$ function contains a ``zero mode'' $\delta (x)$ term 
 \begin{align}
 F_2^{(1)} \left(x,  q^{2}\right) =
& \frac{\alpha_{s}}{\pi} C_{A} q^{2}  \left [ (1-x)\theta (0\leq x\leq 1)-\frac12 \delta( x) \right ]
+\{ x \to -x \} \  . 
\label{xi03f}
\end{align}
However, the function  $F_2^{(1)} \left(x,  q^{2}\right)$ 
disappears in the $q^2\to 0$ limit, and  does not contribute to the ``forward'' PDF $F(x)$.

 Recall that so far ``$x$'' had the meaning of the fraction of initial momentum $p$. 
 So, if the initial momentum is $yp$, then the active gluon  momentum is $xyp$.
 The usual convention is to use $xp$ for the active gluon  momentum and $(x/z)p$ for the
 initial momentum. 
 The ratio of the active gluon momentum to the initial  one is then given by $z$.
 This allows us to use the  kernel given by Eq. (\ref{evo1}) (changing there  $x$ into $z$) to write 
 the evolution equation for the ``forward''  PDF $F(x,\mu^2)$. 
 Since    $F(x, \mu^2)= F(-x, \mu^2)$,  it is sufficient to 
 write the equation  for $x\geq 0$:
   \begin{align}
F (x,\mu_1^2) =&F(x, \mu_2^2) +\frac{\alpha_s}{2\pi}  \,  
\ln \left (\frac{\mu_1^2 }{\mu_2^2}  \right ) \, 
 \int_x^1 \frac{dz}{z}  P_{gg}^F(z) \,  F(x/z, \mu_2^2)
%\nn   &
% \bigg\}
% \nn &
\  , 
\label{evo}
\end{align}
where 
   \begin{align}
P_{gg}^F(z) = 
 \frac{\beta_0}{2} \delta (1-z) + C_A   \bigg [\frac{2}{1-z}\bigg ]_+
%\nn   &
% \bigg\}
% \nn &
\ 
\end{align}
is the $gg$-component of the 1-loop evolution kernel for the ``gluon condensate'' PDF $F(x,\mu^2)$.
The lower limit of integration over $z$ in Eq. (\ref{evo}) reflects the fact that $F(y,\mu^2)$ vanishes for $y>1$.

    \section{Summary and conclusions}

In this paper, we have presented the results for   one-loop corrections to 
the ``gluon condensate'' PDF $F(x)$ introduced in Ref. \cite{Hatta:2020iin}.
In the same paper it was suggested that this twist-4 distribution may have ``zero mode''
$\delta (x)$ terms.  Such terms, in fact,  have been observed in one-loop %perturbative 
diagrams for quark  twist-3 PDFs \cite{Burkardt:2001iy,Aslan:2018tff}.

According to our results, the   $\delta (x)$  terms are absent  in the 
one-loop expressions for the twist-4 gluon PDF $F(x)$. 
Still, we found the $\delta (x)$ term in the one-loop correction 
to the $\xi=0$  GPD $F(x,q^2)$, where $q$ is the momentum transfer
between the initial and the final gluons. This term   is accompanied by the $q^2$ factor, 
and disappears in the forward $q=0$ limit. Another observation is that the spin structure of this
``zero mode'' term is different from that of the tree-level contribution. 
Using the Schwinger parametric representation in our studies  of the ``zero mode''  terms, we have presented a simple way
 of analyzing the origin of  the $\delta (x)$ terms in the Feynman diagram contributions. 
 
 Our  calculations also shed light on   the perturbative evolution properties of 
 the twist-4 ``gluon condensate''  PDF $F(x)$. In particular, the final   result  of our paper provides     explicit   expression for   the  previously unknown $gg$-component 
of its evolution  kernel .

Setting  the   framework for our calculations,   we have proposed to switch to 
  nonforward kinematics. 
Using  this approach, we were able   to get  a nonzero result  for the tree-level matrix element
in a situation when the external gluons are on-shell.  

 Our study has clearly demonstrated the  crucial role  
of  strict compliance with the gauge invariance requirements.  
 Using on-shell external gluons, we have obtained the same  result both in Feynman and light-cone gauges.
This outcome  may be  contrasted  with  the calculations involving  off-shell external gluons,  that have resulted  in  two  different expressions 
 in these two gauges. 
 Attracting attention to  this issue, we  repeat      that our results   do not agree    with 
  those  of the  original  \mbox{attempt \cite{Hatta:2020iin}}  of  one-loop calculations of $F(x)$  for the gluon target. 
  The reason for the discrepancy is the use of virtual external gluons in Ref. \cite{Hatta:2020iin} . 
  
These  observations  create  an alarming  warning 
for  ongoing projects   (see, e.g.,  Ref. \cite{Wang:2019tgg}) to renormalize 
gluon operators on the lattice with the help of    matrix elements calculated
for highly virtual gluon states.

\section*{ Acknowledgements}

This work is supported by Jefferson Science Associates,
 LLC under  U.S. DOE Contract \#DE-AC05-06OR23177
 and by U.S. DOE Grant \#DE-FG02-97ER41028.

 \bibliography{GT4.bib}

\providecommand{\href}[2]{#2}\begingroup\raggedright\begin{thebibliography}{10}

\bibitem{Feynman:1973xc}
R.~P. Feynman, \emph{{\it Photon-hadron interactions}}.
\newblock Reading, 1972.

\bibitem{Ji:2020baz}
X.~Ji, \emph{{Fundamental Properties of the Proton in Light-Front Zero Modes}},
  \href{https://doi.org/10.1016/j.nuclphysb.2020.115181}{\emph{Nucl. Phys.}
  {\bfseries B} (2020) 115181},
  [\href{https://arxiv.org/abs/2003.04478}{{\ttfamily 2003.04478}}].

\bibitem{Hatta:2020iin}
Y.~Hatta and Y.~Zhao, \emph{{Parton distribution function for the gluon
  condensate}}, \href{https://doi.org/10.1103/PhysRevD.102.034004}{\emph{Phys.
  Rev. D} {\bfseries 102} (2020) 034004},
  [\href{https://arxiv.org/abs/2006.02798}{{\ttfamily 2006.02798}}].

\bibitem{Burkardt:2001iy}
M.~Burkardt and Y.~Koike, \emph{{Violation of sum rules for twist three parton
  distributions in QCD}},
  \href{https://doi.org/10.1016/S0550-3213(02)00263-8}{\emph{Nucl. Phys. B}
  {\bfseries 632} (2002) 311--329},
  [\href{https://arxiv.org/abs/hep-ph/0111343}{{\ttfamily hep-ph/0111343}}].

\bibitem{Leibbrandt:1987qv}
G.~Leibbrandt, \emph{{Introduction to Noncovariant Gauges}},
  \href{https://doi.org/10.1103/RevModPhys.59.1067}{\emph{Rev. Mod. Phys.}
  {\bfseries 59} (1987) 1067}.

\bibitem{Aslan:2018tff}
F.~Aslan and M.~Burkardt, \emph{{Singularities in Twist-3 Quark
  Distributions}},
  \href{https://doi.org/10.1103/PhysRevD.101.016010}{\emph{Phys. Rev. D}
  {\bfseries 101} (2020) 016010},
  [\href{https://arxiv.org/abs/1811.00938}{{\ttfamily 1811.00938}}].

\bibitem{Hatta:2020ltd}
Y.~Hatta, \emph{{$CP$-odd gluonic operators in QCD spin physics}},
  \href{https://doi.org/10.1103/PhysRevD.102.094004}{\emph{Phys. Rev. D}
  {\bfseries 102} (2020) 094004},
  [\href{https://arxiv.org/abs/2009.03657}{{\ttfamily 2009.03657}}].

\bibitem{Radyushkin:1983wh}
A.~V. Radyushkin, \emph{{On Spectral Properties of Parton Correlation Functions
  and Multiparton Wave Functions}},
  \href{https://doi.org/10.1016/0370-2693(83)91116-4}{\emph{Phys. Lett. B}
  {\bfseries 131} (1983) 179--182}.

\bibitem{Radyushkin:2017lvu}
A.~V. Radyushkin, \emph{{Quark pseudodistributions at short distances}},
  \href{https://doi.org/10.1016/j.physletb.2018.04.023}{\emph{Phys. Lett. B}
  {\bfseries 781} (2018) 433--442},
  [\href{https://arxiv.org/abs/1710.08813}{{\ttfamily 1710.08813}}].

\bibitem{Balitsky:2019krf}
I.~Balitsky, W.~Morris and A.~Radyushkin, \emph{{Gluon Pseudo-Distributions at
  Short Distances: Forward Case}},
  \href{https://doi.org/10.1016/j.physletb.2020.135621}{\emph{Phys. Lett. B}
  {\bfseries 808} (2020) 135621},
  [\href{https://arxiv.org/abs/1910.13963}{{\ttfamily 1910.13963}}].

\bibitem{Zhao}
S.~Zhao, \emph{{in preparation}}, .

\bibitem{Liu:2019urm}
Y.-S. Liu, W.~Wang, J.~Xu, Q.-A. Zhang, J.-H. Zhang, S.~Zhao et~al.,
  \emph{{Matching generalized parton quasidistributions in the RI/MOM scheme}},
  \href{https://doi.org/10.1103/PhysRevD.100.034006}{\emph{Phys. Rev. D}
  {\bfseries 100} (2019) 034006},
  [\href{https://arxiv.org/abs/1902.00307}{{\ttfamily 1902.00307}}].

\bibitem{Ji:2015qla}
X.~Ji, A.~Sch\"afer, X.~Xiong and J.-H. Zhang, \emph{{One-Loop Matching for
  Generalized Parton Distributions}},
  \href{https://doi.org/10.1103/PhysRevD.92.014039}{\emph{Phys. Rev. D}
  {\bfseries 92} (2015) 014039},
  [\href{https://arxiv.org/abs/1506.00248}{{\ttfamily 1506.00248}}].

\bibitem{Bhattacharya:2020jfj}
S.~Bhattacharya, K.~Cichy, M.~Constantinou, A.~Metz, A.~Scapellato and
  F.~Steffens, \emph{{The role of zero-mode contributions in the matching for
  the twist-3 PDFs $e(x)$ and $h_{L}(x)$}},
  \href{https://doi.org/10.1103/PhysRevD.102.114025}{\emph{Phys. Rev. D}
  {\bfseries 102} (2020) 114025},
  [\href{https://arxiv.org/abs/2006.12347}{{\ttfamily 2006.12347}}].

\bibitem{Chanowitz:1972vd}
M.~S. Chanowitz and J.~R. Ellis, \emph{{Canonical Anomalies and Broken Scale
  Invariance}}, \href{https://doi.org/10.1016/0370-2693(72)90829-5}{\emph{Phys.
  Lett. B} {\bfseries 40} (1972) 397--400}.

\bibitem{Crewther:1972kn}
R.~J. Crewther, \emph{{Nonperturbative evaluation of the anomalies in
  low-energy theorems}},
  \href{https://doi.org/10.1103/PhysRevLett.28.1421}{\emph{Phys. Rev. Lett.}
  {\bfseries 28} (1972) 1421}.

\bibitem{Collins:1976yq}
J.~C. Collins, A.~Duncan and S.~D. Joglekar, \emph{{Trace and Dilatation
  Anomalies in Gauge Theories}},
  \href{https://doi.org/10.1103/PhysRevD.16.438}{\emph{Phys. Rev. D} {\bfseries
  16} (1977) 438--449}.

\bibitem{Kluberg-Stern:1975ebk}
H.~Kluberg-Stern and J.~B. Zuber, \emph{{Renormalization of Nonabelian Gauge
  Theories in a Background Field Gauge. 2. Gauge Invariant Operators}},
  \href{https://doi.org/10.1103/PhysRevD.12.3159}{\emph{Phys. Rev. D}
  {\bfseries 12} (1975) 3159--3180}.

\bibitem{Nielsen:1977sy}
N.~K. Nielsen, \emph{{The Energy Momentum Tensor in a Nonabelian Quark Gluon
  Theory}}, \href{https://doi.org/10.1016/0550-3213(77)90040-2}{\emph{Nucl.
  Phys. B} {\bfseries 120} (1977) 212--220}.

\bibitem{Tarrach:1981bi}
R.~Tarrach, \emph{{The renormalization of FF}},
  \href{https://doi.org/10.1016/0550-3213(82)90301-7}{\emph{Nucl. Phys. B}
  {\bfseries 196} (1982) 45--61}.

\bibitem{Wang:2019tgg}
W.~Wang, J.-H. Zhang, S.~Zhao and R.~Zhu, \emph{{Complete matching for
  quasidistribution functions in large momentum effective theory}},
  \href{https://doi.org/10.1103/PhysRevD.100.074509}{\emph{Phys. Rev. D}
  {\bfseries 100} (2019) 074509},
  [\href{https://arxiv.org/abs/1904.00978}{{\ttfamily 1904.00978}}].

\end{thebibliography}\endgroup
\bibliographystyle{jhep}
 
 \end{document}